\documentstyle[12pt]{article}
\newfont{\bb}{msbm10 scaled \magstep1}
\begin{document}
\begin{titlepage}
\title{
{\bf Discrete Symmetries, Strong CP Problem and Gravity
\footnote{Invited talk at the 4th
Hellenic School on Elementary  Particle Physics, Corfu,
Greece, September 1992.}}
}
{\bf
\author{
Goran Senjanovi\'c \\
International Centre\\
for
Theoretical Physics\\ Trieste, Italy\\
}}
\date{}
\maketitle
\begin{abstract}
\noindent

Spontaneous breaking of parity or time reversal
invariance offers a solution to the strong CP problem,
the stability of which under quantum gravitational
effects provides an upper limit on the scale of symmetry
breaking. Even more important, these Planck scale effects
may provide a simple and natural way out of the resulting
domain wall problem.
\end{abstract}
\thispagestyle{empty}
\end{titlepage}
\pagestyle{empty}

{\large{\bf I. Introduction}}

\vspace{.5cm}
One of the main characteristics of black holes is the
fact that in their vicinity one loses any notion of
global charge \cite{wald:gr}.
The trouble is that if you come close enough to learn
about it, you yourself will be trapped and not able to
send any information out. This of course is not true of
local charges, since through the information carried by
gauge fields one can know about them  even far
from the black hole. In other words, using Gauss
theorem
\begin{equation}\oint \vec E\cdot d\vec S = q\end{equation}
one can obtain the information about the charge $q$ of
the black hole by measuring the electric field away from
it.
It is therefore highly suggestive that the non-gauged
global symmetries, both continuous and discrete are
violated in the presence of black holes \cite{barb:pl}
If so, this should have a profound impact on anything
which utilizes the idea of a global symmetry. Normally,
even if the renormalizable part of the Lagrangian is
invariant under such a symmetry, there is no reason to
expect the same of higher degree effective operators. In
other words, we would expect $d\geq 5$ operators to be cut off
by $1/M_{Pl}$ to break the global symmetry \cite{barb:pl}.
So what, you will say, for any physical effect due to
such terms would be of order $(E/M_{Pl})^n\ (n\geq 1)$,
where $E$ is the relevant energy in the physical process.
For $E\simeq M_{W}$ which is of our interest, this effect is
less than $10^{-17}$! Can this have any impact on present
day physics?

The answer is yes, as I discuss here. I will focus on the
issue of discrete symmetries and the strong CP problem.
The most important role that gravity may play is in
providing a way out of a domain wall problem. Normally, a
spontaneous breaking of a discrete symmetry causes the
formation of domain walls, whose existence is a serious
problem for the big-bang scenario \cite{zeld:jetp}
(unless the scale of symmetry breaking is below 10-100
MeV).
However, an explicit breaking of such a symmetry induced
by quantum gravitational effects (virtual black holes,
wormholes), would provide a source of instability of
domain walls allowing them to dissappear before ever
dominating the energy content of the
universe. This is particularly important in
the case of time reversal $T$ and parity invariance $P$,
symmetries which cannot be gauged (at least not in four
dimensions) \cite{choi:pre}. It is aesthetically
appealing to have these symmetries broken spontaneously,
as suggested in the case of $T (CP)$ more than
twenty years ago by Lee \cite{lee:pr}. It would be only
appropriate that gravity, a theory of space-time dynamics,
may provide a way out of breaking of space-time
symmetries such as $T$ and $P$. Now, there is more to
this than just aesthetics. Spontaneously broken theories
are in general softer, less divergent than the explicitly
broken ones. A spontaneous breaking of either $P$ or $T$
symmetry may thus render the measure of CP breaking by
QCD (through instanton effects) finite and
calculable\cite{bey:prl}.
In other words, in such theories $\theta_{QCD}\leq
10^{-10}$ may be achieved naturally, without any
fine-tuning. We will show here that the gravity effects
(or better Planck scale effects) which explicitly break
CP keep this picture valid as long as the scale of $T$
(or $P$) breaking $\Lambda_T(\Lambda_P)$ is limited from
above \cite{bere:pre}. Thus, these effects may provide a
badly needed upper limit on $\Lambda_T$($\Lambda_P)$, and
offer hope of experimental verification of these ideas.

On the other hand, the most elegant and popular solution
of the strong CP problem is based on the global
Peccei-Quinn axial $U(1)_{PQ}$ symmetry \cite{pecc:prl},
whose breaking leads to the existence of an almost
massless, pseudo Goldstone boson, the axion
\cite{wein:prl}. It is known that the scale of symmetry
breaking $\Lambda_{PQ}$ must be extremely large,
$\Lambda_{PQ}\geq 10^9$ GeV in order for axions not to be
overproduced in stars \cite{kim:pr}. The trouble then is
that the Planck scale effects, if they do break global
symmetries, would spoil this picture completely, leading
in general to a heavy axion and a large $\theta_{QCD}$
\cite{barb:pl}.

Thus quantum gravity may turn out to be a curse for a
Peccei-Quinn scenario and a blessing for the idea of $P$
or $T$ providing a solution to the strong CP problem.

\vspace{1cm}

{\large{\bf II. Discrete Symmetries and a Domain Wall
Problem: Gravity as a Way Out}}

\vspace{.5cm}

In order to illustrate our ideas, let us imagine a simple
case of a discrete symmetry, a real scalar field with no
cubic self-interaction

\begin{equation}
{\cal L}={1\over 2} \partial_\mu\phi
\partial^\mu\phi-{\lambda\over 2} (\phi^2-v^2)^2
\end{equation}
The discrete symmetry $\phi\to -\phi$ is broken
spontaneously through $\langle\phi\rangle =\pm v$. It is
easy to show that the theory allows for a static
classical solution in say $x-y$ plane (the so called kink)

\begin{equation}
\phi_{cl}(z) =v \tanh \left(vz \sqrt \lambda  \right)
\end{equation}
which interpolates between the two physically distinct
values $+v$ and $-v$, since
$\phi_{cl}(z)\mathop{\to}\limits_{z\to\pm\infty} \pm
v$.
This is the infamous domain wall, which carries energy
density per unit area

\begin{equation}
\rho_{cl}(z) =v^3 \left[\coth \left( vz \sqrt \lambda  \right) \right]^4
\end{equation}
expected on physical grounds, since it costs energy to
move from one vacuum to another. It differs from its
vacuum values in a region of width $\delta\simeq
(v \sqrt \lambda)^{-1}$.

In a classic paper Zeldovich et al. have pointed out a
serious cosmological problem imposed by the existence of
these classical solutions. In discussing this problem we
follow here  closely the discussion of Ref. [4],
including also gravity as a potential cure of this
problem.

     Now, at high temperatures
the potential $V({\phi})$ receives an additional
contribution

\begin{equation}
\delta V = {\lambda \over 4}T^2{\phi}^2
\end{equation}

Since ${\delta}V$ is necessarily positive, for sufficiently high
temperature $T > T_c {\approx} v $ the symmetry is restored [12].
In the standard big-bang cosmological scenario, the field
 ${\phi}$
is expected to undergo a phase transition as the universe
cools down from $T > T_c$ to $T < T_c$ .
For separations larger than the correlation length or
horizon size around the time of phase transition, the field
${\phi}$ will independently take either of its vacuum
values giving rise to corresponding domains and domain
walls. To understand the generic features of this
system of domain walls
one may consider the following idealised problem.
Imagine splitting space into cubes of the size of
correlation length. And, say, the probability for the
field to take a particular vacuum value in a given cube
is $p$ ( $= {1 \over N }$, where N is the number of
degenerate vacua ). The nature of domain structure
obtained, a domain being a set of connected cubes
carrying same vacuum value of the field, is a basic
question in percolation theory [13]. The main result we
will need here is that if $p$ is greater than a certain
critical value ${p_c}$, then apart from finite size
domains there will be one and only one
domain of ``infinite'' size formed. For $p$ less than
${p_c}$ there will not be any infinite size domain.
 Generically, i.e. considering even other lattices
than just cubic, ${p_c}$ happens to be less than 0.5 .
In the example of real scalar field we have been
considering, $p$ equals 0.5 .
Thus there would be an ``infinite'' domain
corresponding to each vacuum and therefore infinite
wall of a very complicated topology.
Of course, there will also be a network of finite
size walls.
The question one is interested
in is the energy density contribution of this domain
wall system as it evolves. Following
crude analysis addresses this question [14] .

     The dynamics of the wall is mainly decided by the
force per unit area ${f_T}$ due to tension and
frictional force $f_F$ with surrounding medium. Since
tension in the wall is proportional to the energy
per unit area ${\sigma}$, we get
${f_T \sim}$ ${\sigma \over R}$ for radius of
curvature scale R. Moreover, $f_F {\sim sT^4}$ where
$s$ is the speed of the wall and $T$ the temperature
of the system [15]. When the speed of the wall
has stabilised we have

\begin{equation}
sT^4 = {\sigma \over R}
\end{equation}

\noindent Thus, the typical
time $t_R \sim R / s$
taken by a wall portion of radius scale $R$
to straighten out would be

\begin{equation}
t_R \sim  {R^2T^4 \over \sigma} \approx {R^2 \over {G \sigma t^2}}
\end{equation}

\noindent Making the plausible assumption that
if ${t_R < t }$, the wall curvature on the
scale $R$ would be smoothened out by time t, we
get that the scale on which the wall is smooth
grows as

\begin{equation}
R(t) \approx (G \sigma)^{1\over 2} t^{3 \over 2}
\end{equation}

\noindent Energy density contribution $\rho_W$
to the universe by walls goes as

\begin{equation}
{\rho_W} \sim {{\sigma R^2} \over R^3 } \sim ({\sigma \over
{Gt^3}})^{1\over 2}
 \end{equation}

\noindent Therefore ${\rho_W}$ becomes
comparable to the energy density ${\rho} \sim {1\over {Gt^2}}$ of the
universe in the radiation dominated
era around
${t_0} \sim {1 \over {G\sigma}}$. Thus domain walls
would significantly alter the evolution of the
universe after ${t_0}$.

     Now, the discrete symmetries relevant
 for particle physics
typically tend to be broken at mass scales above
the weak scale ${M_W \approx 100 GeV }$, giving
$t_0 \leq {10^8}$ sec.. This would be
certainly true
of P and T (CP), the examples we are most
interested
in. Hence from above considerations one
would conclude
that discrete symmetries cannot be broken
spontaneously.

      There are two possible ways out of this impasse.
One possibility is that, even for low scales of symmetry
breaking, the phase transition that would have restored the symmetry
does not
take place, at least not until high enough temperatures
to allow inflation to dilute the energy density in the
domain walls. This in general requires a more complicated
Higgs structure than the minimal one
and realistic examples have been discussed in the
literature [16].

 Another way out [14], is the possibility that a spontaneously
broken discrete
 symmetry is also explicitly broken by a small
amount, which lifts the degeneracy of the two vacua
$+v$ and $-v$. For instance, in our example we could imagine
adding to the Lagrangian a small ${\phi}^3$
term which, obviously,
breaks ${\phi} \rightarrow -{\phi}$ symmetry.
 It should not come as a
surprise that this effect may provide a mechanism for
the decay of domain walls; after all now there is a
unique vacuum. Crudely, the way it works is as
follows [14]. Lifting of the degeneracy of the two
vacua by an amount ${\epsilon}$ gives a pressure
difference of the same amount, between
the two sides of the wall, with
a tendency to push the wall into false vacuum region.
Thus the dynamics of the wall is now going to be
decided by combination of the pressure ${\epsilon}$
, forces $f_T$ due to
tension and $f_F$ due to friction mentioned before.
Clearly at some point the forces due to friction and
tension become small, compared to pressure difference
${\epsilon}$, because they are
proportional to $T^4 \sim {1\over Gt^2}$ and
${\sigma \over R} \sim ({\sigma \over {Gt^3}})^{1\over 2}$
respectively. At that
stage the pressure difference will dominate and
cause shrinking of the false vacuum. Actually it is
difficult to find out precisely when the false
vacuum region, and hence the domain walls, disappear.
However, it may be crudely estimated to be the time
when the pressure ${\epsilon}$ exceeds
the force due to tension, or when it exceeds the force
due to friction for a relativistically moving
wall so as to dominate the dynamics. For either
requirement to be satisfied before
$t_0 \sim { 1 \over {G \sigma} } $, the time
when wall contribution $\rho_W$ would
have become comparable to the energy density
of the universe, one obtains

\begin{equation}
\epsilon \geq G{\sigma}^2  \sim  { {v^6} \over {{M_{Pl}}^2}}
\end{equation}

      Of course, it is not very
appealing to introduce ad hoc
the symmetry breaking terms just in order to eliminate
the domain-wall problem. Ideally,
we would prefer these effects
 to be a natural consequence of
underlying theory. An interesting example recently
discussed in the literature
[17] is that of a discrete symmetry
explicitly broken due to instanton induced effects.

\vspace{1cm}

{\bf Role of Gravity}

\vspace{.5cm}

     In this paper we invoke the possibility that the needed
mechanism for explicit breaking
may be naturally provided by gravity. One
expects that gravity, because of black-hole physics,
may not respect global symmetries, both continuous
and discrete ones.
This expectation is motivated by two important points:
 firstly, the
``no-hair" theorems of black-hole physics that state that
stationary black-holes are completely characterised by quantum
numbers
associated with long-range gauge fields, and secondly,
that the Hawking radiation in evaporation of
black-hole is thermal [1]. Now, consider a process in
which a certain amount of normal matter, which is in a
state that is ``odd" under the discrete symmetry in
consideration, collapses under gravity to form a
black-hole. Because of no hair being associated to
the global discrete symmetry, any information
regarding it is lost to observers outside the black
hole. Hawking radiation from the black hole
being thermal in nature does not carry any
information about internal states of the black-hole
either. Of course, it is not certain what the
properties of evaporation would be at late stages
when semi-classical approximation breaks down.
Unless for some reason the processes at late stages
cause the final system to have same global
discrete charges as those of the initial normal
matter that collapsed, the symmetry would stand
violated.

     We wish to note that from very different
viewpoints there have been discussions in the
literature regarding possibility of CP or T violation
in context of gravity. Ashtekar et. al. have
discussed [18] CP "problem" in the framework of
canonical quantisation of gravity.
In Ashtekar variables
reformulation of general relativity, the canonical
variables of the theory resemble those of Yang-Mills
theory. This allows for discussion of ${\theta}$
sectors in canonical quantisation framework for
Yang-Mills to be taken over to the gravity case.
 Moreover, an analogue
of ${\theta FF^d}$ term in the action can also be
given.

      Another set of observations that interest us
particularly were made by Penrose about
T-asymmetry [19]. He contends, based on arguments
related to Bekenstein-Hawking formula,
that there must be some as yet unknown theory
 of quantum gravity that is
time-asymmetric.
 We recall
here only an easy to state, interesting point from
his discussion.
Corresponding to a solution of Einstein's
equation describing collapse of normal matter to
form a black-hole that stays for ever (classically)
, there would be a time-reversed solution,
white-hole, describing
explosion of a singularity into normal matter.
Now, according to Bekenstein-Hawking formula
the surface area $A$, of a black hole's
horizon is proportional to its intrinsic
entropy, $S$

\begin{equation}
S = {kc^3 \over {4\hbar G}} A
\end{equation}

\noindent In classical processes area is
non-decreasing with time and hence so is the entropy.
If an intrinsic entropy is associated with a
white-hole, it is again expected to be proportional to
the area of its horizon. Time
reverse of area principle would give that this
area, and hence the corresponding entropy
can never increase, an anti-thermodynamic behaviour.
 Especially it would be
a strongly
anti-thermodynamic behaviour by the white hole
when it ejects substantial amount of matter.
This is among the reasons that lead Penrose to
consider the possibility that there may be a
general principle that rules out the existence
of white holes and would therefore be time-asymmetric.

     With the premise, in view of preceeding
 discussion, that gravity may violate a
global discrete symmetry we wish to explore
 its consequences for the domain-wall problem.

     The crucial issue one faces in
implementing
this kind of approach is determination of the
precise form of these symmetry-breaking terms. At the
present day understanding of gravity it does not seem
possible to give a satisfactory answer to this question.
The strategy followed in the literature [2], which we also
adopt here,
in analogous discussions
 has been to write
all the higher dimensional effective operators allowed
by gauge invariance of an underlying theory. Of course,
one could take a point of view that the dimension
four and lower terms may also break the discrete
symmetry. We take no stand on this point. In any
case, even if this happens it can only help in
destabilising the domain walls due to increased
symmetry breaking. Our point is that even the tiny
higher dimensional symmetry breaking terms,
cut-off by powers of Planck-mass, may be sufficient
in solving the domain wall problem.

  To illustrate how this works, we turn again to our
simple example of a real scalar field. The effective
higher dimensional operators would take the form

\begin{equation}
V_{eff} = {C_5 \over {M_{Pl}}}{\phi}^5
+ {C_6 \over {M_{Pl}^2}}{\phi}^6 + ....
\end{equation}

     Obviously, all the terms with odd powers of
 ${\phi}$ break the discrete symmetry ${\phi} \rightarrow -{\phi}$.
[We should mention that
while discussing difficulties with a certain
compactification scheme in superstring theory, it
was remarked by Ellis et. al. [20] that a specific
discrete symmetry in their model may be broken by
terms inversely proportional to ${M_{Pl}}$.
But they note further
that massless modes of string theory
would not induce such
terms and that the massive modes, could be the
only possible source of such effects. However, as
we have been pursuing here, the non-perturbative
effects are expected
to be a natural source of breaking
of global discrete symmetries, independent of whether
string theory turns out to be a correct theory of
gravity. Furthermore, as we have emphasised before,
we feel that these effects should be taken
seriously as a possible solution to the domain-wall
problem associated with the fundamental discrete
symmetries of nature such as Parity or Time-reversal
invariance.]

     In estimating precisely the amount of
symmetry-breaking we would need to know the values
of coefficients $C_n$. Barring some
unexpected conspiracy, in the following we will assume
that $C_n$ will be O(1) as they are dimensionless. Moreover,
it is understood that the scale of spontaneous
symmetry-breaking $v$ lies below the Planck scale.
With all this in mind, the
energy-density split between the two vacua would be
$\approx {C_5 \over M_{Pl}}v^5$.
 This is obviously much bigger than the amount
 $v^6 / M_{Pl}^2$ needed to make the domain walls disappear.
This holds true as long as $C_5 \gg v / M_{Pl}$,
which for lower scales of symmetry breaking gets
to be more and more plausible. For example,
if $v = M_{GUT} \approx 10^{15}$ Gev, we need $C_5 > 10^{-4}$,
 whereas for
$v = M_w \approx 100$ Gev, we only need $C_5 > 10^{-17}$ !
 In general, if a leading operator in eq.(10) is of dimension n,
the condition for disappearance of domain walls is
 $C_n > (M_{Pl} / v)^{n - 6}$. Clearly $n=6$ is the critical
value, since for $n>6~C_n$ would have to be unreasonably large
whereas for $n=6~C_6{\sim}O(1)$ may suffice.

\vspace{1cm}

{\large{\bf III.\quad T\quad and \quad P\quad and a
Strong\quad CP\quad Problem}}

 \vspace{.5cm}

As we mentioned in the Introduction, the most popular
solution to the strong CP problem is based on the idea of
$U(1)_{PQ}$ chiral global symmetry. Recall that the
strong CP problem is the fact that the $d=4$
renormalizable part of QCD Lagrangian in general contains
a term
\begin{equation}\Delta{\cal L}_{QCD}=\theta_{QCD}
g^2/32\pi^2\in_{\mu\nu\alpha\beta}
\vec F^{\mu\nu}\vec F^{\alpha\beta}\end{equation}

Although this is a total derivative, due to instanton
effects it does contribute to the action. Now in QCD
alone we could always choose $\theta_{QCD}=0$, but in
the presence of CP violating weak interactions the
effective $\theta$ parameter becomes

\begin{equation}
\bar\theta=\theta_{QCD}+\theta_{QFD}
\end{equation}

where $\theta_{QFD} =\arg\det Mq$ and $Mq$ is the quark mass
matrix. Since $\theta_{QFD}$ is perturbatively
renormalized, setting $\bar\theta=0$ is plagued by
infinities. In the Peccei-Quinn scenario, $\bar\theta$ is
promoted to a dynamical variable, an axion field $a$,
i.e. $\bar\theta=\langle a\rangle$. Due to $U(1)_{PQ}$
breaking by instantons, $a$ is not a purely massless
Goldstone boson, but gets a mass through a potential
$\Delta C\simeq \Lambda^4_{QCD}(1-\cos a/\Lambda_{PQ})$
In other words, $\bar\theta=\langle a/\Lambda_{PQ}\rangle = 0$, but

\begin{equation}
m_a\simeq \Lambda^2_{QCD}/\Lambda_{PQ}
\end{equation}
This is what makes Peccei-Quinn solution to the strong CP
problem so appealing, the fact that $\bar\theta=0$ is a
dynamical phenomenon.

What about Planck scale effect? After all $U(1)_{PQ}$ is
a global symmetry and so we expect the quantum
gravitational effects to break it, i.e. there should be
Planck scale corrections to $\Delta V$
\begin{equation}\Delta V_{Pl}\simeq \Lambda^5_{PQ}/M_{Pl} \cos
(a/\Lambda_{PQ}+\delta)
\end{equation}
where $\delta$ is some arbitrary phase.

Since $\Lambda_{PQ}\geq 10^9$ GeV, the above form
dominates the instanton contribution by at least 26
orders of magnitude, implying $\bar\theta=-\delta$. In
other words, Peccei-Quinn scenario is rendered completely
inoperative (notice that for $\Lambda_{PQ}\leq M_{W}$ it
would still work).

On the other hand, there is another possible strategy
that could provide $\bar\theta$ calculable and small in
the perturbation theory, i.e. $T$ or $P$ symmetry
\cite{bey:prl}. Simply $T$ or (and) $P$ symmetry allows us
to set $\bar\theta=0$. To see how it works, let us take
an example of parity ${\bb P}$, which can be used to set
$\theta_{QCD}=0$. Furthermore, $P$ becomes L-R symmetry
in the fermionic sector, i.e.
$q_L\mathop{\leftrightarrow}\limits^P q_R$ and so
the mass matrix $M_q$ defined through $\bar q_L M_q
q_R$ becomes hermitean: $M_q=M^{\dagger}_q$. This in turn implies
$\theta_{QFD} = \arg\det M_q = 0$, or in other words
$\bar\theta=0$.

The question is what happens next when parity is broken.
Clearly, $\bar\theta$ cannot remain zero, but is
necessarily finite and calculable in perturbation
theory. It's value becomes model  dependant and by now
various models have been offered based either on $P$ or
$T$ symmetry (notice that $T$ can be as effective as $P$)
which keep $\bar\theta$ small, less than it's limit of
$10^{-10}$ \cite{nel:pl}. To see what happens once gravity effects are
switched on we must turn to a specific model. For
simplicity, we discuss first a model of Barr et al.
(called BCS hereafter) \cite{barr:prl} which is based on
a generalized notion of parity, one that transforms
$SU(2)_L\times U(1)$ into its mirror world
$\widehat{SU(2)_L\times U(1)}$. Imagine thus a gauge group

\begin{equation}
G = SU(2)_L\times U(1)\times \widehat{SU(2)_L\times
U(1)}
\end{equation}

with the fermionic assignment $q,\ell$ and $\hat q ,
\hat\ell$ where $q$ and $\ell$ are the ordinary quarks
and leptons, are $\hat q$ and $\hat\ell$ their parity
counterparts. The Higgs sector is the minimal one, i.e.
the usual $SU(2)_L\times U(1)$ doublet $\phi$ and its
parity analog $\hat\phi$. Due to L-R symmetry
$\bar\theta =\theta_q +\theta_{\hat q}$ is forced to
vanish.
When $P$ is broken through $\langle\hat\phi\rangle \gg
\langle\phi\rangle$, $\bar\theta$ is turned on, but due
to the low energy theory being precisely the standard
model, its value must be as small as the finite
 value $\bar\theta$ in the standard model (keep in mind that both
$\langle\phi\rangle $ and $\langle\hat\phi \rangle$ are
real)
\begin{equation}\bar\theta\leq 10^{-19}
\end{equation}
In order to avoid problems with fractionally charged
bound states of ordinary and mirror quarks, one imagines
$U(1)\times \hat U(1)$ symmetry being broken to a
single $U(1)$ at some high enough energy (at the same
this provides a necessary source needed to split the
$\langle\phi\rangle$ and $\langle\hat\phi\rangle$).

Notice the end result of this scenario. Since $M_{\hat
q} ={\langle\hat\phi\rangle\over\langle\phi\rangle}
M^{\dagger}_q$, all the mirror fermion masses are scaled by
the same factor
 \begin{equation}R\equiv
\langle\hat\phi\rangle/\langle\phi\rangle
\end{equation}
compared to the ordinary ones. In other words,
\begin{equation}m_{\hat u}=R m_u,\ m_{\hat e} = R me, {\rm etc.}
\end{equation}
\vspace{1cm}

{\large{\bf IV.\quad Gravity  Effects}}
\vspace{.5cm}

Let us now switch on  higher dimensional Planck
scale effects. Since the model under discussion is not
automatically invariant under $P$, i.e. since $P$ does
not follow from gauge invariance, one could argue perhaps
that even the renormalizable terms in the low energy
theory should be allowed to break $P$. In the absence of
any detailed calculations of the non-perturbative
gravitational effects, one is not able to settle this
issue definitely. In what follows (as in Ref. [8]), we
make an assumption that only Planck scale induced
non-renormalizable terms are allowed, terms which have a
desirable property of vanishing in the $ M_{Pl} \to \infty$
limit. This would be certainly true in any theory with an
automatic $P$ symmetry, a situation we would like to
achieve eventually.

The leading $P$ violating operators are then of
dimension six
\begin{equation}
{\cal L}_{Pl}=\alpha/M^2_{Pl}\bar q_L \phi
q_R \hat \phi^{\dagger}\hat\phi +\ {\rm other\ terms}
\end{equation}
where other terms are assumed to break L-R symmetry and
$\alpha$ is in general a complex matrix. Clearly these
terms do not restrict $\langle\hat\phi\rangle$ severely,
due to the suppression $\langle
\hat\phi\rangle^2/M^2_{Pl}$. Furthermore, being of order
6, they are on the borderline of being large enough to
destabilize the domain walls produced due to the
spontaneous breaking of parity.

The situation becomes more interesting in the case of
$SU(2)_L\times SU(2)_R\times U(1)_{B-L}$ symmetry, as in
the work of Babu and Mohapatra\cite{babu:pr} (which
preceeded the BCS work). The quark mass matrix now takes
the form
\begin{equation}\bordermatrix{&q_R & Q_R\cr
q_L&0 & \Gamma v_L\cr
Q_L &\Gamma^{\dagger} v_R & M\cr}
\label{matrix}
\end{equation}
 where $q_L$ and $q_R$ are
$SU(2)_L$ and $SU(L)_R$ doublets respectively; $Q_{L,R}$
are singlets, $v_L$ and $v_R$ are vev's of $\phi$ and
$\hat\phi$ respectively (called $\phi_L$ and $\phi_R$
) and $\Gamma$ are Yukawa couplings. Since $v_L$,
$v_R\in R$, once again det $M_q=\det\Gamma^{\dagger}\Gamma v_L
v_R$ and thus $\bar\theta = 0$ at the tree level. Babu and
Mohapatra proceed to show that $\bar\theta$ is induced at
the two-loop level and it turns out to be small:
$\bar\theta\leq 10^{-12}$.

In this case the $P$ breaking terms induced by gravity
are of dimension five
\begin{equation}{\cal L}_{Pl}=\beta/M_{Pl} \bar q_L\phi_L\phi^{\dagger}_R q_R+\
{\rm h.c.}
\end{equation}
where $\beta$ is a complex matrix. This means that the
zero in the quark mass matrix (\ref{matrix}) is now induced to be of
order of $\beta v_Lv_R/M_{Pl}$. It can be shown that the
effective $\bar\theta$ induced through this explicit
breaking of $P$ is
\begin{equation}\bar\theta\simeq \beta{v_Lv_R\over m_u
M_{Pl}}\end{equation}
where the leading contribution comes from the up quark
$u$. In order to comply with $\bar\theta\leq 10^{-19}$
one obtains an upper limit on $v_R$, $v_R\leq 10^6$ GeV.

Furthermore, now we can argue that there is no domain
wall problem, since ${\cal L}_{Pl}$ is of dimension five. Thus,
a reasonably low lying left-right symmetry, with
$M_R\leq 10^6$ GeV may account naturally for the smallness
of $\bar\theta$ without running into conflict with either
experiment or cosmological considerations.

Alternatively, one can utilize CP instead of parity. A nice example is
provided in \cite{berez}, which is based on the horizontal family
symmetry. Similarly, as before, when the gravity effects are switched on
one obtains an upper limit on the horizontal symmetry breaking scale
of approximately $10^6$ GeV.

\vspace{1cm}

{\large{\bf V. Summary and Discussion}}
\vspace{0.5cm}

Understanding quantum gravity remains one of the central
issues of today's theoretical physics and before it is
achieved in depth, it is important to look for the probes
of its potential low energy effects. One promising
direction lies in searching for the possible violations
of global symmetries, expected through the experience
with black holes. The notion of global symmetries, if
valid at all, may be subject to breaking terms in the
effective theory at low energies, cut-off by the Planck
scale.
On the other hand, the solutions to the strong CP problem
which  try to explain naturally the smallness of strong
CP parameter $\bar\theta\leq 10^{-9}$ are in general
based on global symmetries. In particular, Peccei-Quinn
mechanism is based on the $U(1)_{PQ}$ global symmetry and
seems to be unable to survive the gravitational effects.
On the other hand, the approach which utilizes discrete
symmetries, such as parity or time-reversal invariance
tend to benefit from such effects, since they can provide
an upper limit on the scale of breaking of $P$ or $T$.
All this assuming that the Planck scale phenomena vanish
in the limit $M_{Pl}\to\infty$.

Finally, in my opinion the most interesting aspect of
these Planck scale effects lies in the possibility of
resolving the domain wall problem. The symmetry breaking of
the order of $\langle\phi\rangle/M_{Pl}$ (or even
$\langle \phi\rangle^2/M_{Pl}$) is enough to cause the
dissappearance of domain walls before they dominate the
energy density of the universe. In other words, the idea
of spontaneous breaking of discrete symmetries may be
free from cosmological difficulties and moreover it may
play (through the explicit breaking by gravity) a role in
baryogenesis.

\end{document}